\documentclass[12pt]{article} 
\usepackage{feynmp}
\catcode`\@=11 
 
\def\mycite{\@ifnextchar [{\@tempswatrue\@mycitex}{\@tempswafalse\@mycitex[]}} 
\def\mcite{\@ifnextchar [{\@tempswatrue\@mycitex}{\@tempswafalse\@mycitex[]}} 
 
\def\@mycitex[#1]#2{\if@filesw\immediate\write\@auxout{\string\citation{#2}}\fi 
 \def\@citea{}\@mycite{\@for\@citeb:=#2\do 
    {\@citea\def\@citea{,\penalty\@m\ }\@ifundefined 
       {b@\@citeb}{{\bf ?}\@warning 
       {Citation `\@citeb' on page \thepage \space undefined}}%
\hbox{\csname b@\@citeb\endcsname}}}{#1}}

\def\@mycite#1{[{#1}]} 
 
\catcode`\@=12 
\def \ov {\over}
\def \foot{\footnote}
\def \bi {\bibitem}

\def \ci{\cite}
\def \a {\alpha}
\def \del {\partial}

\newcommand{\rf}[1]{(\ref{#1})}
\newcommand{\non}{\nonumber \\*}
\newcommand{\be}{\begin{equation}}
\newcommand{\ee}{\end{equation}}
\def\bea{\begin{eqnarray}}
\def\eea{\end{eqnarray}}

\def\la{\left\langle}
\def\ra{\right\rangle}
\def\d{\partial}
\newcommand\N{${\cal N}=4~$}
\newcommand\Tr{\mathop{\mathrm{Tr}}}

\newcommand\Sp{\mathop{\mathrm{Sp}}}
\renewcommand\Im{\mathop{\mathrm{Im}}}
\newcommand\sign{\mathop{\mathrm{sgn}}}
\def\D{\delta}
\def \m {\mu}
\def \n{\nu}

\def\ep{\varepsilon}
\def\l{\lambda}

\def\wz{S_{\rm WZ}}
\def\gm11{\Gamma^{11}} 
\def \NN {{\cal N}}

\begin{document} 

\title{
\vspace{-0.5cm}
\hfill{\small OHSTPY-HEP-T-99-028}\\
\vspace{0.1cm}
Magnetic Interactions of D-branes \\ \vspace{-0.2cm}
and Wess-Zumino Terms \\ \vspace{-0.2cm}
in Super Yang-Mills Effective Actions \\ 
}
\vspace{-3cm}
\author{\large 
A.A. Tseytlin$^{a,}$\thanks{Also at Lebedev 
Physics Institute, Moscow and Imperial College, London.}
\mbox{} and \mbox{} K. Zarembo$^{b,c,}$\thanks{Also at Institute 
of Theoretical and Experimental Physics, Moscow.}
\vspace{0.1cm} \\ \small  
$^a$Department of Physics 
\vspace{-0.3cm} \mbox{} \\ \small 
The Ohio State University
\vspace{-0.3cm} \mbox{} \\ \small 
Columbus, OH 43210-1106, USA
\mbox{}
\vspace{-0.1cm} \\ \small  
$^b$Department of Physics and Astronomy 
\vspace{-0.3cm} \mbox{} \\ \small 
$^c$Pacific Institute for the Mathematical Sciences
\vspace{-0.3cm} \mbox{} \\ \small 
University of British Columbia
\vspace{-0.3cm} \mbox{} \\ \small 
Vancouver, BC V6T 1Z1,  Canada
\mbox{}
}
\date {}
\maketitle
\vspace{-1cm}
\begin{abstract}
\vspace{-0.2cm}
\baselineskip=12pt
There is a close relation between classical supergravity
and quantum SYM descriptions of interactions
between separated branes. In the case of D3 branes 
the equivalence of leading-order potentials is due to
non-renormalization of the  $F^4$  term in $\NN=4$  SYM theory. 
Here we point out the existence of another special
non-renormalized term in quantum SYM effective action.
This term reproduces the interaction potential  between
electric charge of a D3-brane probe and magnetic
charge of a D3-brane source,  represented  by the 
Chern-Simons part of the D-brane action. This 
unique Wess-Zumino term depends on all six scalar
fields and originates from a phase of the 
euclidean fermion determinant in SYM theory.
It is manifestly scale invariant (i.e. is the same 
for large and small separations between branes) 
and can not receive higher loop corrections in gauge
theory.  Maximally supersymmetric  SYM theories
in $D=p+1 > 4$ contain  mixed WZ terms which depend
on both scalar and gauge field backgrounds,
and  which reproduce the corresponding CS terms in the 
supergravity interaction potentials between separated 
Dp-branes for $p >3$.  Purely  scalar WZ terms 
appear in other cases, e.g., in half less supersymmetric
 gauge theories 
in various dimensions  describing  magnetic
interactions between D$p$ and D$(6-p)$  branes.
\end{abstract}

\newpage\setcounter{page}{1}
\setcounter{equation}{0} 

\section{Introduction}

The duality between the supergravity and the world-volume descriptions
of D-branes has led to important advances in 
 understanding of
dynamics of supersymmetric gauge theories.
Many aspects of interactions of D-branes
which are transparent
in the supergravity description, 
translate into quite non-trivial
properties of  the world-volume field theory. 
The electric and 
gravitational interactions 
between branes have been widely studied
 in this context (see, e.g., \ci{DOU,BA,TA,LIF,CT,MAL}
and references there).
 The agreement between the supergravity 
and the SYM descriptions
of leading-order interactions between branes 
 can be traced 
to the universal non-renormalization properties 
of the ${ F}^4$ terms in the effective action of
 maximally supersymmetric gauge 
theories in various dimensions \ci{DS,SP}.

We shall discuss magnetic interactions.
The  prime case   of interest in connection with 4-d gauge
theories
is a D3 brane moving in the background of 
other D3 branes.
The self-duality of the RR  five-form field strength implies
that a D3 brane
carries both electric and magnetic charges. As a
consequence, 
the probe brane
will experience the Lorentz force, similar
 to the one an
electric charge  
experiences in the  magnetic field of a monopole.
For example,  the action of a  D3-brane probe 
moving in a  supergravity background  produced 
by a D3-brane source contains \ci{LI,DO,BGM}
the Chern-Simons term $\int_4 C_4 =\int_5 F_5$  
which describes  the interaction of an 
electric  charge of the probe with the  electric and the 
magnetic fields produced by the source. In the static gauge, 
the electric interaction produces the  effective  ``-1" contribution to  
 the  P-even Born-Infeld 
part of the 
D3-brane action 
$S= S_{\rm BI} + S_{\rm SC}^{(el)}+ S_{\rm SC}^{(mag)}$\ \foot{Here 
 \ci{POL} 
 $T_3 = {1 \ov 2 \pi g_s }, \
  Q= { 1 \ov \pi}  N g_s $, \ $2 \pi \a'=1$, \ 
   $|X|^2 \equiv 
X^I X^I$. }
\be 
  S_{\rm BI} +  S_{\rm SC}^{(el)}  = - T_3  \int d^{4}  x \ 
   {|X|^4 \ov Q} \bigg[ \sqrt {-\det \left(  \eta_{\m\n}  + 
{Q\ov  |X|^{4} }\del_\m X^I \del_\n X^I  + 
 {Q^{1/2} \ov |X|^{2}} F_{\m\n} \right)} -1\bigg]  . 
 \label{BI}\ee
 This term ensures  the vanishing of the interaction 
potential between static parallel branes. 
The  magnetic interaction term (which is real
 for  Minkowski  signature)
$S_{\rm CS} =  N  \wz 
\sim  i N \int_{5}  
 \epsilon_{I_1 ...I_6} { 1 \ov |X|^6}   X^{I_1}
  d
X^{I_2} \wedge ...\wedge d X^{I_6}$
is non-vanishing only when all 6 scalars  have non-trivial
gradients.

Since  many features of the magnetic D3 brane
 interaction
are 
similar to those  of the  Lorentz  interaction 
between   an electric charge  
 and  a  magnetic monopole,  
let us briefly review some facts about the latter case. 
The Lorentz force acting on the charge 
in the field of   magnetic monopole is 
\be
{\cal F}^i=q_e\,\ep^{ijk}\,\,\frac{q_m}{4\pi
|X|^3}\,X_j\dot{X}_k
\ .   \ee
The Dirac string singularity of the gauge potential
for  
the monopole field does not allow the Lorentz force to  
be a 
variation of a well-defined local action functional. 
The variational principle for a 
charged particle interacting with a  monopole can be formulated only
by adding the {\it non-local}
 Wess-Zumino term to the action
\bea
\wz&=&-\frac{q_eq_m}{8\pi}
\int d^2\tau \,\ep^{ab}\,\ep^{ijk}\,\,\frac{1}{|X|^3}\,
X_i \d_a X_j \d_b X_k 
\non &=&
-\frac{q_eq_m}{8\pi}
\int \,\ep^{ijk}\,
n_i dn_j \wedge dn_k,~~~~~\ \ \ \   n_i\equiv \frac{X_i}{|X|}\
. 
\eea
Here the integration is over a domain whose boundary is the
time axis.
Then the  variation of the WZ action 
reproduces the Lorentz force:
 $\D \wz/\D X_i={\cal F}^i$. 

According to the standard argument, 
there is an ambiguity in the definition of the 
WZ action, because $X_i$
is defined at  the boundary of the integration
 domain and can be continued
into the interior in an arbitrary way. Since the integrand is
locally
a total derivative, this ambiguity is discrete, in the sense
that
the values of the action for different continuations differ
 by
an integer multiple of $ q_eq_m$.  
Despite the multi-valuedness of the action, the  path
integral  remains single-valued,  provided  
$q_eq_m$ is an integer 
multiple of $2\pi$. The consistency of the quantum
mechanics of an electrically  charged particle 
 thus leads to the Dirac
quantization condition for the magnetic charge: 
$q_m=2\pi n/q_e$.

The Lorentz force experienced by a  D3 brane
 in the background of another
D3 brane  (with both branes having unit charges)  is 
\be\label{lf}
{\cal
F}^I=\frac{1}{12\pi^2}\,\ep^{\mu\nu\l\rho}\,\ep^{IJKLMN}\,\,\frac{1}{|X|^6}
X_J\d_\mu X_K
\d_\nu X_L \d_\l X_M \d_\rho X_N,
\ee
$$ \m,\n= 0,1,2,3\ , \ \ \ \ \    I,J,...= 1,2,3,4,5,6 \ . 
$$
Here $X_I(x)$ parametrize the position of the probe brane in
the 
6-d space transverse to the source brane as a function of 
the 4
longitudinal
coordinates $x^\mu$.
The Lorentz force 
can be represented as a variation of the five-dimensional
integral ($m= 0,1,...,5$):\foot{The WZ term 
can be
written also as a 4-d  integral of  a local  
functional by using spherical $S^5$ coordinates
instead of Cartesian $X^I$ (see \ci{KAM}).}
\be
{\cal F}^I=\frac{\D \wz}{\D X_I}\ ,
\ee
\bea
\wz&=&-\frac{1}{60\pi^2}\,\int d^5x\,\ep^{mnklr}
\ep^{IJKLMN}\,\,\frac{1}{|X|^6}\,
X_I \d_m X_J 
\d_n X_K \d_k X_L \d_l X_M \d_r X_N
\non
&=&-\frac{1}{60\pi^2}\,\int\ep^{IJKLMN}\,
n_I dn_J \wedge
dn_K \wedge dn_L \wedge dn_M \wedge
dn_N\ , 
\label{six}
\eea
where $ n_I=\frac{X_I}{|X|}$ parametrize $S^5=SO(6)/SO(5)$.
This action is defined up to $2\pi$, which reflects the fact
that a
D3 brane carries one unit of the quantized magnetic charge.

This ``topological"  term is scale-invariant 
and 
 does not depend on gauge coupling. 
 Its coefficient cannot be renormalized 
 since any  non-trivial dependence  on the dilaton would 
 break gauge invariance.  Also, as  for many 
  other WZ terms, 
 the  renormalization of its
 coefficient  would   contradict the
  topological magnetic  charge quantization condition.\foot{
 Assuming  that there are no phase transitions in the
coupling constant, the
coefficient  should be  a smooth function
of the coupling,  but   an  integer valued
smooth function must be  a constant.} 
Like the one-loop ${ F}^4$ term, this WZ term  should  
    thus  be a rare example
  of an exact non-renormalized SYM 
  interaction.\foot{The two terms 
   may, in fact, be related by 
  a  non-linear part of  maximal $D=10$
   supersymmetry 
  as is suggested by the existence of the 
     action for a 
   D3-brane  in $AdS_5 \times S^5$  space constructed in 
  \ci{MT}, where the BI and CS terms in the  action
  are related by the $\kappa$-symmetry.
  Also, there seems to be  a close analogy
   with the 
  ${ F}^4$  and $\epsilon_{10}  { F}^4 C_2$ terms
   in the   type I superstring 1-loop effective 
   action  which are related by $D=10$ supersymmetry
   and obey a  non-renormalization 
   theorem \ci{TTT}.}

  The scale ($X \to a X$) invariance and the topological 
  nature of the  above  WZ term suggests that 
  it should  follow from   the string-theory 
  description  of D-brane interaction
  at both large (supergravity) and small (gauge theory)
  distance regions.
 The non-renormalization of the coefficient of this 
 term implies that  it should be present 
 in the effective action  of 
  {\it quantum}  \N  $SU(N)$ SYM theory
  on the
 Coulomb
branch of the  moduli space  in both 
weak   and strong  coupling regions.

Therefore, it  should  be 
  expected, 
 both from the weakly coupled string theory
 ``long-distance--short-distance" duality
  \ci{DOU} and the AdS/CFT duality \ci{MA}, 
 that,  
 like the P-even ``${ F}^4/X^4+$superpartners" term  
 in $S_{\rm
 BI}$, this P-odd term 
  should  be exactly
 reproduced by the 1-loop computation 
 on the  \N SYM  theory side.

Our aim below  is  to confirm 
this expectation 
 by the  explicit computation of the  imaginary part 
 of the fermion determinant in the SYM theory. 
 As far as we are aware, the derivation of this 
 $D=4$  WZ term  from  \N SYM  theory 
  was 
missing in the literature
(the calculations presented in  \cite{bth,gm}, 
though   similar, led to a different class of 
CS terms, see below). 
That this term should  have, by analogy with the
case
discussed in  \ci{BD},  
  a Berry phase 
interpretation  was suggested to one of us (A.T.)
 by M. Douglas (for some related work 
 in the context of matrix models see also
 \ci{IT}). 
 
 Our direct perturbative 
 derivation  of the WZ term \rf{six}
 in the effective action of \N SYM
theory
  described 
  below in Section 2 will not be referring 
 to the Berry phase.
 We will  show that
the WZ term 
originate from the same hexagon
 diagram which 
is responsible
for the  chiral anomaly 
in ten-dimensional SYM theory \cite{hexagon}.

In Section 3 we will describe  a 
similar computation  of  counterparts 
of the ten-dimensional anomaly graph 
in other supersymmetric gauge theories
describing systems of separated Dp branes with $p >3$.
In contrast to the $D=4$ SYM case 
(and some of its analogs discussed below) 
where 
the WZ term is {\it purely scalar}  and does
not have a local representation, 
the  hexagon graphs  in $D > 4$ lead 
to a {\it different}   class of   
WZ terms which involve both scalars  {\it and}
 vectors  and admit a {\it local}  CS-type representation.
 This class of WZ terms was  previously 
derived  from SYM theories
in \cite{bth,gm}. 


Section 4 will contain some concluding remarks.

\section{Wess-Zumino term in the $D=4$  \N SYM theory}

The WZ term is odd in time derivatives, 
so it  has a factor of $i$ in the Euclidean action
(below we  choose  the Euclidean signature of  the metric).
  The effective action induced by the  bosonic SYM 
  degrees
of freedom is real, so the only potential source
of the imaginary WZ  term  is the 
(gauge-invariant, $O(6)$ invariant and
conformal-invariant) 
phase of the fermion determinant. The appearance of 
a WZ term 
in the phase
of a  fermion determinant is not unusual, 
 and examples of chiral WZW
terms induced by fermionic loop  are known
 \ci{Dia83,DH}. 
 The present case
of scalar fields interacting with fermions 
in \N  $D=4$ SYM theory  was  not explicitly 
 discussed before.

The  part of the 1-loop
 effective action in \N SYM theory
which is induced by the  4 Weyl fermions 
has the following form 
(after continuation to 
Euclidean space) 
\be
S_{\rm ferm}=-\frac{1}{2}\,\Tr \ln \left[
\left(\Gamma^0\Gamma^\mu\d_\mu+i\Gamma^0\Gamma^I
[\Phi_I,\ \cdot]\right) 
\frac{1+\gm11}{2}\right] \ . 
\ee
Here $\Gamma^M$ are ten-dimensional Dirac matrices:
\be
\{\Gamma^M,\Gamma^N\}=2\D^{MN}\ ,~~~~~
(\Gamma^{M})^\dagger=\Gamma^M\ , \ \ \ \ \ \ 
\gm11\equiv i\Gamma^0\ldots\Gamma^9\ .\ee
We  assume that the  10-d indices are split in the 4+6 way, 
$M=(\mu, I)$,  and that the $D=4$  vector field has trivial
background.
Generalization  to the case of non-trivial $A_\m$ 
background in $D >4$ is
 straightforward (see \cite{bth,gm} and  Section 3 below), 
 but in the 
 case of \N SYM theory the
  WZ term happens to  depend only on the
   six   scalar fields.

The system  of $N$ ``slowly moving"
 separated D3 branes  is represented by 
 slowly varying diagonal scalar fields:
\be\label{scal}
\Phi_I=\left(
\begin{array}{ccc}
X^1_I&&\\
&\ddots&\\
&&X^N_I
\end{array}
\right).
\ee
The commutator term in the Dirac operator in this background vanishes 
for diagonal components of fermions, 
 and for non-diagonal it  becomes
$[\Phi_I,\Psi]^{ab}=(X_I^a-X_I^b)\Psi^{ab}$. The effective action
thus decomposes  into  a sum of pairwise interactions:
$$
S_{\rm ferm}=\sum_{a<b}S(X^a-X^b)\ ,
$$
where
\be
S(X)=-\Tr\left[\ln\left(\Gamma^0\Gamma^\mu\d_\mu+i\Gamma^0\Gamma^I X_I\right)
\frac{1+\gm11}{2}\right]\,.
\label{teen}
\ee
Taking the variation, we find
($\Sp$ is the trace in spinor indices):
\be
\frac{\D S}{\D X_I(x)}=\Sp\left[\la x \left|\frac{1}{i\Gamma^\mu\d_\mu
-\Gamma^JX_J}\right| x \ra
\Gamma^I\,\frac{1+\gm11}{2}\right]\,.
\ee
We are interested in the {\it  imaginary}
 part of the effective action. Since
the operator $i\Gamma^\mu\d_\mu-\Gamma^JX_J$ is Hermitean,
 and 
$\Gamma^I(1+\gm11)
=(1-\gm11)\Gamma^I$, taking the difference of 
the above expression and its complex conjugate
 it is easy to see that 
\be
\frac{\D \Im S}{\D X_I(x)}=\frac{1}{2i}\,
\Sp\left [ \la x \left|\frac{1}{i\Gamma^\mu\d_\mu-\Gamma^JX_J}\right| x\ra
\Gamma^I\,\gm11\right] \,.
\ee
For slowly varying fields, this expression can be expanded in derivatives of 
$X_I$. The term with $n$ derivatives will be proportional 
to  the trace of $2n+3$
Dirac matrices. Since this trace contains $\gm11$,  and
\be
\Sp \left(\Gamma^{M_1}\ldots\Gamma^{M_k}\gm11\right)=0~~
\ \ \  {\rm for}~~k<10\ ,
\ee
the expansion wil start with the {\it four}  derivative term
coming from the diagram
(dashed lines are external scalar fields):

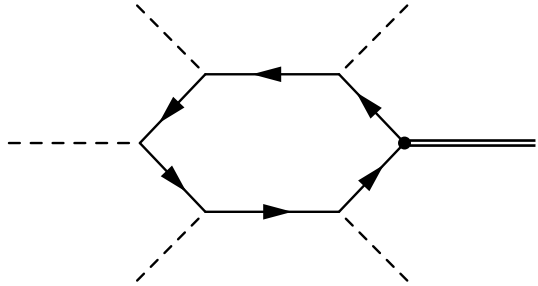
\begin{center}
\begin{fmffile}{loop}
\begin{fmfgraph}(70,40)
\fmfsurround{v1,v2,v3,v4,v5,v6}
\fmf{double}{v1,i1}
\fmf{dashes}{v2,i2}
\fmf{dashes}{v3,i3}
\fmf{dashes}{v4,i4}
\fmf{dashes}{v5,i5}
\fmf{dashes}{v6,i6}
\fmfcyclen{plain_arrow}{i}{6}
\fmfdot{i1}
\end{fmfgraph}
\end{fmffile}
\end{center}
Analytically,
\bea\label{hex}
\frac{\D \Im S}{\D X_I}&=&-\frac{1}{2i}\,
\int\frac{d^4k}{(2\pi)^4}\,
\frac{X_J}{(k^2+X^2)^5}\,\,
\d_\mu X_K \d_\nu X_L \d_\l X_M \d_\rho X_N
\non
&&\times
\Sp\left( \Gamma^\mu \Gamma^K\Gamma^\nu\Gamma^L
\Gamma^\l\Gamma^M\Gamma^\rho\Gamma^N\Gamma^J
\Gamma^I\gm11\right) 
+O(\d^5).
\eea
Using the identity
\be
\Sp \left(
\Gamma^{M_1}\ldots\Gamma^{M_{10}}\gm11\right)
=32i\ep^{M_1\ldots M_{10}}\ , 
\ee
and doing the momentum integral, we find that 
the variation of the
imaginary part of the effective action reproduces 
exactly the Lorentz force between a pair of D3 branes
 \rf{lf}:
\be
\frac{\D \Im S}{\D X_I}=
\frac{1}{12\pi^2}\,\ep^{\mu\nu\l\rho}\ep^{IJKLMN}\,\,\frac{1}{|X|^6}\,
X_J\d_\mu X_K
\d_\nu X_L \d_\l X_M \d_\rho X_N+O(\d^5).
\ee
For  the case of a single D3 brane interacting with 
a cluster of $N$ coinciding D3 branes we get
$S_{\rm ferm} = N S(X)$, and thus rederive  the  magnetic
Chern-Simons term in  the D3 brane probe action
directly from the gauge theory. 

The imaginary part of the SYM effective action contains higher-derivative
corrections which are not seen on the supergravity side. 
The conformal symmetry of \N SYM  implies that 
the extra derivatives are
necessarily accompanied by extra powers of $1/|X|$,
and so 
the higher-derivative terms are no longer invariant
 under the rescaling
$X\rightarrow X/\alpha'$. From the supergravity point 
of view  one 
 may interpret these 
 terms as being effectively of 
  higher order in $\alpha'$. Indeed, 
  the long-distance or closed
  string channel representation  of the full stringy expression 
  for the D3-brane interaction amplitude 
  should contain, in particular, 
    all massless mode (SYM) 
     1-loop corrections of the open string channel.

\section{WZ terms  in $D > 4$ SYM theories }
The $D=4$ case discussed in the previous section 
is special in that the WZ term there depends 
only on the scalar
fields.
An analogous  but actually 
different class 
of  `magnetic' WZ terms appears in the effective 
actions of maximally supersymmetric  SYM theories in 
higher dimensions 
$ 4 <  D=p+1 < 9$ \cite{bth,gm}. 
 For $D>4$ one needs to switch also a
non-trivial gauge field background in order to 
get a non-zero result for the imaginary part of the
fermionic determinant.
This  has a natural interpretation  on the supergravity
side: while D3 branes  carry both electric and magnetic
charges and thus their interaction potential 
contains
`magnetic'  contribution, 
separated  magnetic ($p >3$)
Dp-branes interact only `electrically', 
unless one 
switches  on a gauge field
background which 
induces   effective  electric charges on the 
Dp brane probe.

For example, 
the action of a D5 brane probe 
moving in the D5 brane background contains 
the  CS term ($I=1,2,3,4$) 
\be
 S_{\rm CS}\  \propto\ 
i \int_6\ C_2 \wedge  F \wedge F\ \propto\ 
i N \int_7  \epsilon_{IJKL} 
{ 1 \ov |X|^4}  X^I dX^J\wedge  dX^K \wedge d X^L
\wedge F \wedge F \ , 
\label{pya}
\ee
where $dC_2$ is the (magnetic) gauge field strength of the
D5 brane source. The corresponding WZ term indeed arises
in the effective action of 
$D=6$ SYM theory describing
multiple D5 branes. 

Let us   consider the general case of the maximal 
SYM theory obtained by dimensional 
 reduction of $D=10$ SYM 
theory to $D=p+1 <   10$, 
and couple the fermions 
to both the  diagonal 
scalar background  \rf{scal} and the
abelian gauge
field background ($\m=0,1,2, ...,p$) 
\be
A_\mu=\left(
\begin{array}{ccc}
A^1_\mu&&\\
&\ddots&\\
&&A^N_\mu
\end{array}
\right).
\ee
As in the $p=3$ case \rf{teen},  the fermionic contribution
to the 1-loop 
 effective action factorizes
$$
S_{\rm ferm}=\sum_{a<b}S(X^a-X^b,A^a-A^b)\ ,
$$
where now 
\be
S(X,A)=-\Tr\left[\ln\left(\Gamma^0\Gamma^\mu\d_\mu+
i\Gamma^0\Gamma^\mu A_\mu+i\Gamma^0\Gamma^I X_I\right)
\frac{1+\gm11}{2}\right]\,.
\ee
The first term in the derivative expansion 
of the imaginary part of this action
 comes from the
hexagon  diagram
  and
 has the form similar
to eq.~\rf{hex} with $(p+1)$-dimensional momentum integral instead
of 4-dimensional one and with some of the scalar
 fields replaced 
by the gauge potentials.

The contribution of the hexagon diagram in 
various dimensions
$5\leq D\leq 9$ was shown to give rise to 
 {\it local} Chern-Simons terms
in the effective action \cite{bth}. Below we rederive these Chern-Simons
terms and
clarify their relation to the WZ actions.

If $p\leq 7$, we can take a variation with respect to the
scalar field to get:
\bea
\frac{\D\Im S}{\D X_I}
=&-&  (8-p)d_p \ 
\,\ep^{\mu_1\ldots\mu_{p+1}}\ep^{IJK_1\ldots K_{7-p}}\,
\non && \times
\frac{1}{|X|^{9-p}}\,\,X_J\,\d_{\mu_1}X_{K_1}\ldots\d_{\mu_{7-p}}
X_{K_{7-p}}F_{\mu_{8-p}\mu_{9-p}}\ldots
 F_{\mu_p\mu_{p+1}}\  , 
\eea
\be 
d_p \equiv
\frac{(-1)^{\frac{p(p-1)}{2}}}{4(p-3)!(4\pi)^{\frac{p}{2}}
\,\Gamma\left(\frac{10-p}{2}\right)}\ . 
\ee 
As a result, the effective action contains the 
term:
\bea
\Im S&=& d_p 
\int d^{p+2}x\,\ 
\ep^{\mu_0\ldots\mu_{p+1}}\ep^{JK_0\ldots K_{7-p}}
\non && \times 
\frac{1}{|X|^{9-p}}\,\,X_J\,\d_{\mu_0}X_{K_0}\ldots\d_{\mu_{7-p}}
X_{K_{7-p}}F_{\mu_{8-p}\mu_{9-p}}\ldots F_{\mu_p\mu_{p+1}}\
,
\eea
or, equivalently ($n_I = X_I/|X|$) 
\be
\Im S= \ d_p\  2^{p-3}\ 
\int \ep^{I_1\ldots I_{9-p}}\,n_{I_1}\, dn_{I_2}\wedge\ldots
\wedge dn_{I_{9-p}}(\wedge {F})^{p-3}\ .
\label{ter}
\ee
This expression  reduces to our previous $D=4$ SYM result
 \rf{six} in the case of $p=3$. For $p=5$ this WZ term 
  reproduces  
the  CS interaction \rf{pya}
in the supergravity description.

Note that the  nonlocal nature of the action \rf{ter} 
is fake. Since 
$F=dA$,  we can  integrate by parts and that leads 
 to the local Chern-Simons form
given in \cite{bth}.
For example, the local CS form 
of  \rf{pya}
is $\int_6 dC_2(X) \wedge F \wedge A$.
 This `integration by parts'
 is not possible in the case of purely scalar WZ terms.
 
Like the CS terms in the Dp brane actions, the 
 WZ terms with  different $p$ in \rf{ter}
 are related 
by dimensional reduction (`smearing' and T-duality, 
$A_i\to X_i$).

Instead 
    of computing the variation
    of the effective action 
     over $X_I$, 
 another way to  obtain 
  these  `mixed' WZ terms is to calculate
   the 
induced current \cite{bth}
\be
{\cal J}^\mu=\frac{\D \Im S}{\D A_\mu}\ .
\ee
For $p=9$  the divergence of 
this current produces the chiral anomaly
\cite{hexagon},
which makes the  ten-dimensional non-abelian 
 SYM theory inconsistent.
 For $p=8$, the induced current is\foot{If $X=0$ somewhere, 
this current is not
conserved:
$$
\d_\mu {\cal J}^\mu=\frac{1}{12288\pi^4}\,\delta(X)\,
\ep^{\mu_0\mu_1\ldots\mu_8}\,\d_{\mu_0} X 
F_{\mu_1\mu_2}\ldots F_{\mu_7\mu_8}.
$$
The equation $X=0$ determines a domain wall in the nine-dimensional theory,
which  appears in any field configation satisfying
boundary conditions $X\rightarrow \pm X_0$ at $x^1\rightarrow\pm\infty$,
though
these domain walls do not exist as classical solutions, because
there is no potential for $X$.
It is known that domain walls support fermion zero modes independently
of the wall profile. As a consequence, the
effective 8D field theory on the wall contains chiral fermions, which make
the effective theory anomalous. The chiral
anomaly of the fermion zero modes on the wall exactly compensates
the divergence of the current ${\cal J}^\mu$ \cite{ch85}.
The hypersurface $X=0$ corresponds to the intersection of the D8-branes.
The total inflow of the electric charge into the intersection
is equal to the Chern class $c_4(F)\sim \int 
 (\wedge F)^4$, where $F$ is
a (relative) magnetic field on the
 D8 branes.
This is essentially equivalent to the D8--D0 
case discussed in  \ci{DOE, KL},  since the 
magnetic  flux $\int (\wedge F)^4$ on one of the
D8-branes 
represents a  D0-brane charge. Thus 
the above phenomenon is  a  reflection 
  of  string
creation phenomenon when D0 passes through D8 \ci{KL}.
  }   
\be\label{curr}
{\cal J}^\mu=\frac{1}{12288\pi^4}\,\sign ( X) \ 
\,\ep^{\mu\mu_1\ldots\mu_8}F_{\mu_1\mu_2}\ldots
 F_{\mu_7\mu_8}\  , \ \ \ \ \ \ \ 
 \sign (X) = {X_9\ov |X_9|} \ . 
 \label{ooo}
\ee


In \ci{gm} it was suggested that the 
 theory of multiple M5 branes should contain 
 a similar `mixed'  $\int_6 B_2 \wedge  H_4(X) $, \ 
 $H_4(X) = \epsilon_{IJKLM} 
{ 1 \ov |X|^5}  X^I dX^J\wedge  dX^K \wedge d X^L 
\wedge d
X^M$, \ 
CS 
 term related by compactification on $S^1$ 
 to the $ \int_5 A \wedge  H_4(X) $ CS
 term in   $D=5$ SYM (multiple D4 brane)
 theory,  and the one-loop 
  microscopic derivation  (equivalent 
   to the one in 
  \ci{bth}) 
  of the  latter term 
was  given.

Analogous    WZ terms are found  in other
D-brane interaction systems
described  by gauge theories
with less than maximal supersymmetry.
For example, pure-scalar WZ terms appear 
in  the case of Dp---D(6-p)  (electric-magnetic)
brane interaction.  It is  possible  to see
that in the case of the D5--D1 system\foot{Related 
case of  M5--M2  magnetic
interaction was considered in \ci{BD}.
 D6-D0 system
 was discussed 
 in \ci{BIL}.
} 
described by a particular 
 $\NN=4$,  $D=2$ supersymmetric 
gauge theory  \ci{DPS}, 
 the relevant fermion
determinant  contains 
the WZ term 
$\int_3 \epsilon_{ijkl} { 1 \ov |X|^4} X_i dX_j \wedge 
dX_k
\wedge 
dX_l $ which reproduces the 
CS term $\int_2 C_2$  in the action of a D-string 
probe  moving in the magnetic background produced  by 
a D5-brane source. 
The same term can be obtained  
 by starting with the 
D5--D5 system with the WZ term \rf{pya}, and 
compactifying 4 parallel directions on a torus
and assuming that the  gauge  field background
has a non-zero magnetic flux $\int_4 F \wedge F$ 
representing the D1-brane charge.
This is a particular example of (T-duality) relations between 
different magnetic WZ terms in \rf{ter}.

The abelian WZ terms discussed above 
have natural  non-abelian generalizations.
In particular, they should reproduce 
the 
non-abelian CS terms in multiple D-brane action 
given  in \ci{MY}.\foot{Additional 
 [X,X] factors originate from internal components of
 $F_{ij}$, i.e. these CS term may be  related  by 
dimensional reduction and T-duality to 
$\int {\rm tr}(F \wedge ...\wedge F) \wedge C_2 +...$ 
CS terms
in D9 brane action.}

Let us note also that SYM theories 
defined on  {\it curved} $D$-dimensional 
spaces should contain  curvature-dependent WZ terms
similar to \rf{ter} with $F$ replaced by $R$
(and other  mixed terms). They should  
reproduce 
the corresponding $R$-dependent CS terms \ci{ber,BGM,ser}
in the Dp-brane actions 
(for example, $\int_6 R\wedge R \wedge C_2$ in D5 brane
case).

\section{Discussion}

We have shown that the 1-loop effective action of $D=4$ \ 
\N SYM theory contains  the  unique
WZ term \rf{six}  coming from the phase of the Euclidean 
fermion determinant. The presence  of this term 
(and of its $D > 4$ analogs (22))  is related 
to the existence of chiral anomaly  in
 $D=10$ SYM theory.
This  term  depends only on 6 scalar fields, 
is manifestly $SO(6)$ and conformal invariant,   
and  its  coefficient should not be renormalized 
by higher loop  corrections.

One interesting open question is how to construct
a supersymmetric generalization of this term.
The answer does not seem obvious since the use 
of either $\NN=1$ or $\NN=2$ superfield  formulation
of \N SYM theory breaks the $SO(6)$ symmetry. 
The WZ term \rf{six} is actually the integral 
part of the  space-time supersymmetric and 
$\kappa$-symmetric action \ci{MT} 
for a D3 brane  propagating 
in $AdS_5 \times S^5$  vacuum of type IIB supergravity.
As in the similar superstring action case \ci{MMT}, 
this term  must be added to the Born-Infeld part
\rf{BI}  of
 the action  to ensure its 
 $\kappa$-symmetry.  Fixing the static gauge 
 and a  $\kappa$-symmetry  gauge 
 in the action of \ci{MT} in a suitable way
 one should be able to  read off the 4-d 
  supersymmetric
 form of this WZ term.
 After the gauge fixing,    half of the original
 32 superconformal 
 symmetry generators become
  non-linearly realized, and they should be 
 relating the WZ term to the  terms in the BI part
  of the action.
 
 It should be possible to rederive this WZ term 
 directly from string theory, by taking an appropriate
 $\a'\to 0$  limit in the 1-loop expression for the
 interaction potential between two 
 separated D3-branes. The  topological nature of this
 term  suggests that it should be originating from
 certain fermionic zero mode contribution.

 Similar purely scalar magnetic 
  WZ term $\int_3  C_3$ 
  appears in the classical  action of M2 brane moving
 in the background of an M5 brane source. 
 Though  lack of detailed  understanding of the theory of
 multiple
 M-branes  prohibits us from  deriving  
 this term directly from a  microscopic  theory
 (as was possible  in the  D-brane case), 
 it is natural to expect (by analogy  with 
 a related  proposal about a CS term in the  M5 brane
 theory  action    \ci{gm}) 
 that this
WZ term is again universal, i.e.   
 is not renormalized.

The magnetic WZ terms  are present also  in 
some orbifold   theories
\cite{orb}. For example, they appear in 
the field theory on a stack of an equal
number of electric and magnetic 
D3-branes in type 0 string theory
\cite{KlTs}, which is a $Z_2$ orbifold 
of \N SYM \cite{NekShat}.
This theory has two sets  of scalar fields, 
$X_{\rm (el)}$ and
$X_{\rm (mg)}$, which correspond to the
 transverse coordinates of the 
electric and the magnetic branes. The Yukawa 
couplings in the diagonal
background of these scalar fields are 
$\bar{\Psi}^{ab}\Gamma^I(X_{\rm (el)}^a-X_{\rm (mg)}^b)_I\Psi^{ab}$
\cite{TsZa}. 
This Yukawa interaction induces a WZ term 
depending on the difference 
$X_{\rm (el)}^a-X_{\rm (mg)}^b$. 
The same prediction  (CS term in the interaction
potential)
should follow  from the gravity description,
 since there should be a  Lorentz force 
between the separated electric and the magnetic D-branes.

\subsection*{Acknowledgements}
A.A.T. would like to acknowledge
 I. Chepelev for a collaboration at  an 
  initial stage,  
and M. Douglas, S. Kuzenko, R. Metsaev  
 and J. Polchinski for useful discussions and remarks.
The work of A.T. is  supported in part by
the  DOE grant  DOE/ER/01545-780, 
by the EC TMR   grant ERBFMRX-CT96-0045,  by the 
INTAS grant No.96-538 and 
by the NATO grant
 PST.CLG 974965.
The work of K.Z. is supported by  
PIms Postdoctoral Fellowship, NSERC of Canada, RFFI grant
98-01-00327 and grant 96-15-96455 for the promotion of
scientific schools.


\end{document}